\documentclass[aps,twocolumn,superscriptaddress,showpacs]{revtex4}
\usepackage{graphicx}

\begin{document}

\title{Dipolar effect in coherent spin mixing of two atoms in a single optical lattice site}

\author{B.~Sun}
\affiliation{School of Physics, Georgia Institute of Technology,
Atlanta, Georgia 30332, USA.}

\author{W.~X.~Zhang}
\affiliation{Ames Laboratory, Iowa State University, Ames, Iowa
50011, USA.}

\author{S. Yi}
\affiliation{Institute of Theoretical
Physics, The Chinese Academy of Sciences, Beijing 100080, China}

\author{M. S. Chapman}
\affiliation{School of Physics, Georgia Institute of Technology,
Atlanta, Georgia 30332, USA.}

\author{L.~You}
\affiliation{School of Physics, Georgia Institute of Technology,
Atlanta, Georgia 30332, USA.}

\date{\today}

\begin{abstract}
We show that atomic dipolar effects are detectable in the system
that recently demonstrated two-atom coherent spin dynamics within
individual lattice sites of a Mott state. Based on a two-state
approximation for the two-atom internal states and relying on a
variational approach, we have estimated the spin dipolar effect.
Despite the absolute weakness of the dipole-dipole interaction, it
is shown that it leads to experimentally observable effects in the
spin mixing dynamics.
\end{abstract}


\maketitle

The recent successes of coherent spin mixing
\cite{MSZhang,HS,Bloch1} inside Bose condensed atoms, driven by
reversible collisions between pairs of atoms
$|M_F=0\rangle+|M_F'=0\rangle\leftrightarrow
|-1\rangle+|1\rangle$, have generated significant interest in the
quantum dynamics of atomic spins \cite{jason,machida,law,kurn}.
These experiments cover a broad limit from a condensate with large
number of atoms \cite{MSZhang,HS} to many identical trapping sites
as in an optical lattice each containing two atoms \cite{Bloch1},
and raise significant hope for the long discussed applications of
atomic quantum gases to the emerging field of quantum information
science. As observed in these experiments, the spin coherence time
rivals the best motional state coherence ever achieved on neutral
atoms \cite{eric} and is ideally suited for quantum information
processing applications \cite{zoller,Bloch}.

In this paper, we reveal an interesting observation: the atomic
spin dipolar effects seem to be detectable in the system that
recently demonstrated two-atom coherent spin mixing \cite{Bloch1}.
Dipolar interactions are ubiquitous in atomic systems. It offers a
unique playing field because of the two significant differences
from the the nominal $s$-wave short-ranged interactions: it is
anisotropic and long-ranged \cite{dipyou}. These interesting
properties have stimulated intensive research of dipolar
condensates \cite{yi1,goral1,santos,giov,pfau}.


Our model system consists of two spinor atoms ($^{87}$Rb)
inside a harmonic trap. Corrections due to the anharmonic trapping
potential of a standing wave optical
lattice \cite{artur} will be addressed elsewhere.
We will consider the simple case of a cylindrical harmonic trap
instead of the spherical harmonic trap
as in the experiment \cite{Bloch1},
because the dipolar interaction is known to display a sensitive dependence on
the trap aspect ratios \cite{yi1}. The spherical harmonic
trap minimizes the dipolar effect and thus provides an excellent
calibration for the experimental systems,
particularly the small spin exchange interaction \cite{MSZhang,artur}.
Following the usual procedure of separating the total motion
into the center of mass (CM) and the relative (rel) motion,
the system Hamiltonian becomes
$H = H_{\mathrm{CM}} + H_{\mathrm{rel}}$ with
\begin{eqnarray}
H_{\mathrm{CM}} &=& -{\hbar^2 \nabla^2_{\vec{R}}\over 2M} + {1\over
2}M\omega_\rho^2(X^2+Y^2+\lambda^2Z^2),\\
H_{\mathrm{rel}} &=& H_0 + H_s + H_{dd}+ H_{B},
\end{eqnarray}
for the CM- and rel-motion, respectively, with coordinates
$\vec{R} = {(\vec{r}_1 + \vec{r}_2) / 2}$ and
$\vec{r} = \vec{r}_1 -\vec{r}_2 $. $M=2m$ is the total
mass, while $\mu={m/ 2}$ is the reduced mass.
$H_{\mathrm{rel}}$ contains several parts as outlined below,
\begin{eqnarray}
H_0 &=& -{\hbar^2\nabla^2_{\vec{r}}\over 2\mu}  + {1\over 2}\mu\omega_\rho^2(x^2+y^2+\lambda^2z^2), \\
H_s &=&(c_0+c_2\vec{F}_1\cdot\vec{F}_2)\delta^{({\rm reg})}(\vec{r}), \\
H_{dd} &=& {c_d\over r^3} \left[\vec{F}_1\cdot\vec{F}_2
-3(\vec{F}_1\cdot\hat{r})\cdot (\vec{F}_2\cdot\hat{r})\right], \label{dip}
\end{eqnarray}
where $H_0$ describes the rel-motional,
and $H_s$ accounts for the regularized s-wave contact
interactions between two spin-1 atoms with $\delta^{({\rm reg})}(\vec{r})
\equiv \delta(\vec{r})({\partial/
\partial r})r$. $H_{dd}$ denotes the spin dipolar interaction.
The first
order Zeeman effect does not contribute due to a zero
magnetization, while the second order Zeeman effect
$H_{B}\approx \pm 72(\vec F\cdot\vec B)^2$ (Hz/Gauss$^2$)
in reference to the $M_F=0$ state with the `+' and `$-$' signs
for $F=1$ and $F=2$ respectively. The various interaction
coefficients are listed below
\begin{eqnarray}
c_0 &=& {4\pi \hbar^2\over m} {a_0 + 2a_2\over 3}, \\
c_2 &=& {4\pi \hbar^2\over m} {a_2 - a_0\over 3}, \\
c_d &=& {\mu_0\over 4\pi} g_F^2 \mu_B^2.
\end{eqnarray}
$a_{0(2)}$ is the scattering length for the combined channel of total $F=0(2)$.
$g_F$ is the Land$\acute{e}$ g-factor for the hyperfine spin state of $F$,
and $\mu_B$ is the Bohr magneton.

Before presenting our theoretical analysis, we comment on the
strength of spin dipolar interactions \cite{pu}. In a scalar
condensate, the dipolar effect is usually calibrated against the
nominal $s$-wave interaction, i.e., one simply compares $|c_d|$
with $|c_0|$ and uses their ratio as a parameter. This ratio can
be increased by decreasing $c_0$ perhaps through a Feshbach
resonance \cite{jila}. For spin dipoles, however, the spin
exchange interaction coefficient $c_2$ also needs to be compared.
The extremely small $c_2$ for $^{87}$Rb serves to enhance the spin
dipolar effect because the spinor nature of the ground state
energy is determined by $c_2$ not by the much larger $c_0$. Using
$|c_d|/|c_2|$ as a parameter, $^{87}$Rb could be viewed as a
stronger spin dipolar condensate than Cr. Although in a polarized
condensate, Cr atoms $(F=3)$ with a dipole moment of 6$\mu_B$
enhance significantly the dipolar interaction. When the spinor
nature of the dipole is of interest \cite{pu,dipsantos,ueda}, the
dipolar effect for Cr is weaker because of the large spin exchange
interactions \cite{pu,ho}.

We first will estimate the relative strength of different
interaction terms for the situation as in Ref. \cite{Bloch1}. For
an optical lattice $\propto
V_0[\sin^2(kx)+\sin^2(ky)+\lambda^2\sin^2(kz)]$ of a depth $V_0=s
E_r$ in units of the recoil energy $E_r$ for the lattice laser,
each single lattice site is approximated like a harmonic trap with
a radial frequency $\omega_\rho=\sqrt{ 2s E_r k^2 / m}= {2\sqrt{s}
E_r / \hbar}$
and a Gaussian relative motional ground state
\begin{eqnarray}
\phi_0(\vec{r}) = {\lambda^{1/4} \over \pi^{3/4}  a_\rho^{3/2}}\,
\exp\left({-{\rho^2\over 2\,a_\rho^2}- {\lambda z^2\over 2 a_\rho^2} }\right),
\label{gw0}
\end{eqnarray}
with a radial width $a_\rho=\sqrt{\hbar/(\mu\omega_\rho)}$
($\rho=\sqrt{x^2+y^2}$). The center of mass motional ground state
$\Phi_0(\vec R)$ takes the same form except for the width
$A_\rho=\sqrt{\hbar/(2m\omega_\rho)}$. For a wavelength of $840$
nm, we find $\omega_\rho \sim (2\pi) 41,126.3$ Hz, and
$a_\rho=0.0752$ $\mu$m for $s=40$. We further note that $a_0=
101.8a_B $ and $a_2 =100.4a_B$, with $a_B$ being the Bohr radius
\cite{coef}. These give rise to typical density interaction $c_0
\langle n\rangle_0 /\hbar\sim (2\pi)6,589$ Hz, spin exchange
interaction $c_2 \langle n\rangle_0 /\hbar\sim (2\pi)(-30)$ Hz,
and spin dipolar interaction $c_d \langle n\rangle_0/\hbar \sim
(2\pi)2.71$ Hz for the spherically symmetric case of $\lambda=1$,
all much less that the trap level spacing. Thus, a reasonable
estimate of the motional state would be the non-interacting ground
state $\phi_0(\vec{r})$ from which the averaged density becomes
\begin{eqnarray}
\langle n \rangle_0=2\int |\phi_0(\vec r)|^4 d\vec{r}
={\sqrt{\lambda}\over \sqrt{2}\,\pi^{3/2} a_\rho^3}.
\end{eqnarray}
We further note the relative strength of $c_2/c_0 \sim
-0.00462657$ and $c_d/c_2 \sim -0.0902$, which encourages a
Gaussian variational ansatz \cite{yi1}.

The dominant mixing interaction is spin exchange that couples the
two-atom internal state $|M_F=0,M_F'=0\rangle$ to $|1,-1\rangle$.
The spin dipolar interaction is about 5 times smaller and averages
to vanishingly small net effect for a spherical symmetric motional
state. We therefore will limit our discussions to the above two
internal states, a picture uniformly adopted by the
experimentalists \cite{MSZhang,HS,Bloch1}. We have further carried
out numerical simulations at $B=0$ of the full system dynamics,
including all other spin states that are coupled by the dipolar
term, i.e., with the complete Hilbert space of spin degree of
freedom: $|M_F=-1,0,1;M_F'=-1,0,1\rangle$. For a spherical trap
with the same (averaged) frequency, the probability of atoms in
other spin states that do not conserve the total magnetization
($|1,1\rangle$, $|-1,-1\rangle$, $|0,1\rangle$, and
$|0,-1\rangle$), which are responsible for the dipolar relaxation,
is found to be less than $10^{-6}$ in the first oscillation period
of coherent spin mixing; even for a trap with $\lambda=3$, this
probability remains negligible, and is in fact only several times
enhanced. Thus the negligibly small dipolar spin relaxation in
$^{87}$Rb \cite{coef} makes the two-state model an excellent
approximation for our system. At finite values of the B-field,
except for accidental resonances when other spin states in higher
motional states are shifted into near resonance with the two-state
doublet in ground motional state, the linear Zeeman effect
generally leads to large detunings, also validates the
approximation. Even at accidental resonances, the total population
out of the two-state doublet is found to be only $\sim 10^{-3}$
for a spherical trap. For Cr atoms, however, more effort is needed
to understand the conditions for spin mixing dynamics due to the
much enhanced dipolar relaxation \cite{sr}.

The two-atom wave function is then approximated as
\begin{eqnarray}
\alpha_{0,0}|0,0\rangle\psi_{0,0}(\vec r_1,\vec r_2)
+\alpha_{1,-1}|1,-1\rangle\psi_{1,-1}(\vec r_1,\vec r_2). \hskip
12pt
\end{eqnarray}
This leads to a spin mixing matrix element of
\begin{eqnarray}
{1\over 2}
\hbar\Omega = \int d\vec r_1\int\!\!&d\vec r_2&\!\!
\psi_{0,0}^*(\vec r_1,\vec r_2)
\langle 0,0|(H_s\nonumber\\
&& +H_{dd})
|1,-1\rangle\psi_{1,-1}(\vec r_1,\vec r_2). \hskip 18pt
\label{g0}
\end{eqnarray}

As a first approximation, the relative motion is
simply taken to be the ground state of the harmonic trap, i.e.,
$\psi_{M_F,M_F'}(\vec r_1,\vec r_2)=\Phi_0(\vec{R})\phi_{M_F,M_F'}(\vec{r})$
with $\phi_{0,0}(\vec{r})=\phi_{1,-1}(\vec{r})=\phi_{0}(\vec{r})$.
This leads to
\begin{eqnarray}
{1\over 2} \hbar\Omega &=& \int d\vec r
\phi_{0,0}^*(\vec{r})\langle 0,0|H_s+H_{dd}
|1,-1\rangle\phi_{1,-1}(\vec{r}).  \hskip 12pt
\label{feq}
\end{eqnarray}

An improved approximation is the variational calculation, labelled as $\phi^{(v)}$,
for the relative motional state including
the dipolar interaction as have been used extensively in the past \cite{yi1}.
We take a Gaussian ansatz with its widths $w_{\rho/z}$ as variational
parameters \cite{yi1}
$$\phi_{1,-1}(\vec{r}) = {1\over \pi^{3/4}(w_{\rho}^2 w_z)^{1/2}}
\,\mathrm{exp} \left(- {x^2+y^2\over 2 w_{\rho}^2} -
{z^2\over 2 w_z^2} \right),
$$
the relative energy functional
then becomes
\begin{eqnarray}
{E\over \hbar\omega_{\rho}} &=& {1\over 4}\left(2{a_\rho^2 \over
w_{\rho}^2} +{a_\rho^2 \over w_{z}^2}   \right) + {1\over
4}\left(2{w_{\rho}^2 \over a_\rho^2} + \lambda^2 { w_{z}^2 \over
a_{\rho}^2} \right)\nonumber\\
&& + {c_0-c_2\over
\pi^{3/2}w_{\rho}^2w_{z}\hbar\omega_{\rho}}
 -{2\over 3\sqrt{\pi}}{c_d \over
w_{\rho}^2w_{z}\hbar\omega_{\rho}} \chi(\kappa), \hskip 18pt
\label{Ey}
\end{eqnarray}
with $\chi(\kappa)=[2\kappa^2+1-3\kappa^2H(\kappa)]/[
2(\kappa^2-1)] + (\kappa^2-1) H(\kappa)$ and $H(\kappa)=
{\mathrm{tanh}^{-1}\sqrt{1-\kappa^2}/ \sqrt{1-\kappa^2} }$. $a_{z}
= \sqrt{\hbar/ (\mu \omega_{z})}$ is the axial width of the trap,
and $\kappa=w_{\rho}/w_z$ is the aspect ratio of the variational
ground state, non-spherical (or $\kappa\neq 1$) even in a
spherical harmonic trap with $\lambda=\omega_z/\omega_\rho=1$
because of the dipolar interaction \cite{yi1}.
$\phi_{0,0}(\vec{r})$ is obtained from the result of
$\phi_{1,-1}(\vec{r})$ by excluding the dipolar interaction or
taking $c_d=0$ and adjusting to its own $s$-wave scattering
strength by taking $c_2=0$.

A mean field approach is sometimes used in the
literature where the two-atom motional state
$\psi_{M_F,M_F'}(\vec r_1,\vec r_2)$ is approximated by
$\phi_{c}^{M_F,M_F'}(\vec{r}_1)\phi_{c}^{M_F,M_F'}(\vec{r}_2)$
as for a two-atom condensate with
$\phi_{c}^{M_F,M_F'}(\vec{r})$ obtained from the corresponding
Gross-Pitaevskii equation
\begin{eqnarray}
\left[-{\hbar^2\nabla_{\vec{r}}^2\over 2m}+{m\omega_\rho^2\over
2}(\rho^2+\lambda^2 z^2) +V_{\rm
int}\right]\phi_{c}(\vec{r})=\mu_c \phi_{c},
\end{eqnarray}
with $V_{\rm int}=c_0|\phi_{c}(\vec{r})|^2$, or
\begin{eqnarray}
V_{\rm int}&=&(c_0-c_2)|\phi_{c}(\vec{r})|^2\nonumber\\
&&-c_d\int d\vec r'{1\over |\vec r-\vec r'|^3}(1-3\cos^2\theta)|\phi_{c}(\vec{r}')|^2, \hskip 12pt
\end{eqnarray}
respectively, for the internal states $|0,0\rangle$ or
$|1,-1\rangle$. $\mu_c$ is the chemical potential, and $\theta$ is
the angle between the $z$-axis and $\vec r-\vec r'$. A Gaussian
variational approach is proven to be adequate within the parameter
regions of interest \cite{yi1,santos,blume}. Like the $\phi_0$
approximation, atom-atom correlation \cite{ceder} is neglected
because of the use of product motional states here.

We have used the momentum space pseudo-potential \cite{Derevianko}
$v(\vec{k},\vec{k}') = -{\hbar^2 /(2\pi^2
m)}a_{sd}'\sqrt{5}[P_2(\mathrm{cos}\theta_{k'}) + ({k/k'})^2
P_2(\mathrm{cos}\theta_{k}) ]$ with $a_{sd}' = {\sqrt{2}mc_d/(
12\sqrt{5}}{\hbar^2})$ for a more accurate evaluation of the
dipolar term in Eq. (\ref{Ey}), which can differ upto 50\% for the
parameter range reported here. Because the complete motional wave
function is Gaussian shaped, we find that $\langle
\phi_{1,-1}|H_{dd}|\phi_{0,0}\rangle$ can be evaluated
analytically. In the results shown below for two
$^{87}\mathrm{Rb}$ atoms, the spin mixing effective Rabi frequency
is defined as $\Omega_{\rm eff}=\sqrt{\Omega^2+\Delta^2}$, and
$f_{\rm eff}=\Omega_{\rm eff}/2\pi$. $\Delta$ is the bare energy
difference: $\Delta\equiv (E_{1,-1}-E_{0,0})/\hbar$.

\begin{figure}[h] \centering
\includegraphics[width=3.00in]{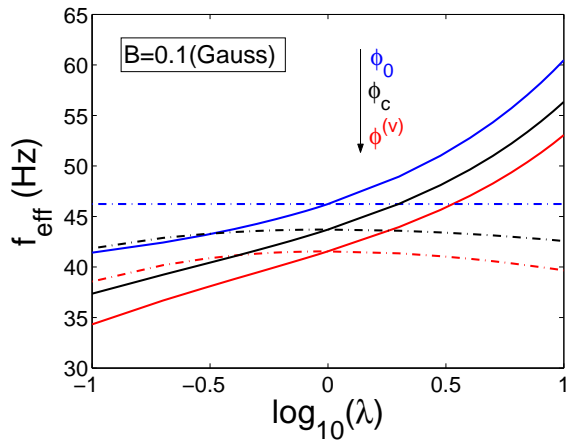}
\caption{(Color online) The aspect ratio dependence of $f_{\rm
eff}$ computed with the three approximation schemes as labelled.
The three solid lines include dipolar interaction, while the
dot-dashed lines are the corresponding ones without dipolar
interaction. Here we fix
$(\omega_\rho^2\omega_z)^{1/3}=(2\pi)41.1$(kHz).} \label{fig1}
\end{figure}

In Fig. \ref{fig1}, we show the dependence of $f_{\rm eff}$ on the
trap aspect ratio $\lambda$ at an external magnetic field of
$B=0.1$ (Gauss). To facilitate a fair comparison, we have fixed
the geometric average of the trap frequencies
$(\omega_\rho^2\omega_z)^{1/3}$ to the spherical trap of
$\omega_{\rho}=\omega_z=(2\pi)41.1$(kHz) \cite{Bloch1}. Somewhat
surprising at first sight is the noticeable quantitative
differences (within experimental sensitivity) among the different
approximations. At $\lambda=1$ for a spherical trap, our results
also differ from the experimental observations \cite{Bloch1}. We
note that the harmonic approximation of $V_0\sin^2(kx)$ by
$m\omega_\rho^2x^2/2$ introduces about a $5\%$ error \cite{artur}.
Although the spread among the different approximation schemes
calls for a more accurate treatment for the motional state, the
dipolar effect due to the $H_{dd}$ term in Eq. (\ref{feq}) is
quite accurately reproduced to less than $1\%$ \cite{note}. An
improved treatment of the relative motional wave function is also
needed inside a cylindrical trap, if spin mixing is used to
calibrate atomic interactions like what has been accomplished for
a spherical harmonic trap \cite{artur}.
\begin{figure}[h]
\centering
\includegraphics[width=3.00in]{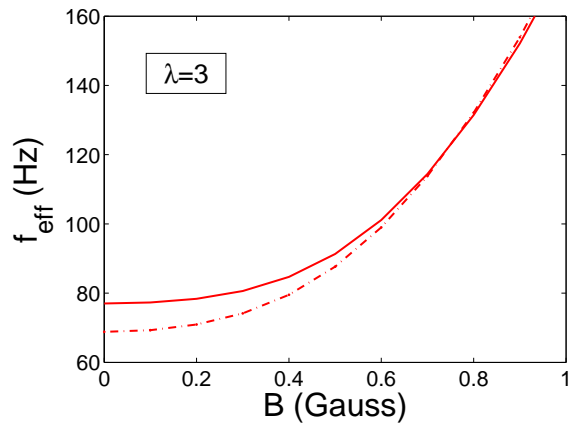}
\caption{(Color online) The B-field dependence of $f_{\rm eff}$
computed within $\phi^{(v)}$ approximation for $\lambda=3$. Solid
line includes dipolar interaction while the dot-dashed line
excludes dipolar interaction. Here, unlike in Fig. \ref{fig1}, we
fix $\omega_{\rho}=(2\pi)41.1$(kHz).} \label{fig2}
\end{figure}

Based on the current experimental sensitivity, dipolar effects
should be detectable for $\lambda>3$ and can constitute a more
than $10\%$ increase in $f_{\rm eff}$. They are minimized for a
spherical trap as shown in Fig. \ref{fig2} with an actual shift of
about $10^{-4}$ or less for the experiment of Ref. \cite{artur}.

\begin{figure}[h]
\centering
\includegraphics[width=3.00in]{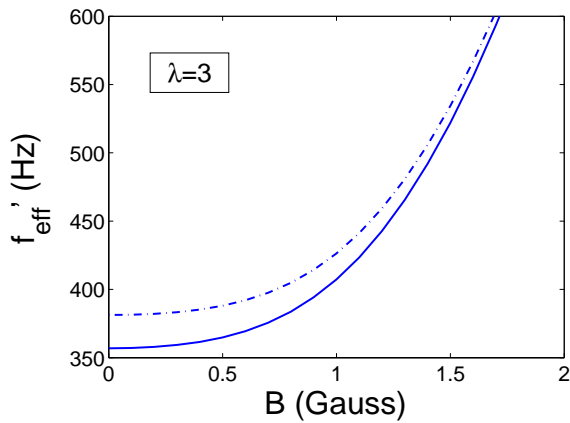}
\caption{(Color online) The same as in Fig. \ref{fig2} except for
spin mixing between $|-1,-1\rangle$ and $|0,-2\rangle$ for $F=2$.
} \label{fig3}
\end{figure}

Also we have studied the $F=2$ manifold \cite{zgm}, where spin
mixing dynamics is generally dominated by three two-atom internal
states $|0,0\rangle$, $|1,-1\rangle$, and $|2,-2\rangle$ at zero
magnetization \cite{Bloch1,artur}.
We find that spin dipolar effect remains observable in the
frequencies of the various mixing channels.
More interesting is the two-state mixing
channel of $|-1,-1\rangle\leftrightarrow |0,-2\rangle$ with a
nonzero magnetization as shown in Fig. \ref{fig3}. Inside an
oblate trap, dipolar interaction is dominantly repulsive in state
$|-1,-1\rangle$ in contrast to the attractive state $|1,-1\rangle$
of the $F=1$ case. At a weak magnetic field and for $\lambda=3$,
we find the dipolar interaction constitutes a 25 Hz downward shift
computed within the $\phi^{(v)}$ approximation.

Dipolar interaction is normally stronger by a factor of two
in magnitude along the direction of the dipoles
(favored by the $\lambda<1$ geometry)
in comparison to the perpendicular direction (favored when $\lambda>1$).
Our results on the spin mixing frequency, however, reveals a
completely opposite trend; we find relatively larger (smaller)
dipolar effects for $\lambda>1$ ($<1$).
This can be easily understood. For the $F=1$ case,
the dipolar term $H_{dd}$ is positive (negative)
for $\lambda<1$ ($>1$), thus destructively (constructively) add to the
(negative) $c_2$ term in Eq. (\ref{feq}). For the $F=2$ case,
the same reasoning applies despite of the opposite dipolar
interaction in state $|-1,-1\rangle$. The dipolar effect becomes
constructively enhanced for $\lambda>1$ because the $c_2$
is positive in $F=2$.

Before conclusion, we note that we have also calculated the spin
mixing frequency for two $^{52}$Cr atoms with $F=3$. In the limit
of very large ($\lambda =10$) and very small aspect ratios
($\lambda =1/10$), dipolar interaction causes about an 8\%
difference in the spin mixing frequency, which is about four times
smaller than in $F=1$ of $^{87}$Rb at $\lambda =10$, and close to
each other at $\lambda =1/10$. Similar conclusions hold for
$^{87}$Rb in $F=2$. It is in this sense we say that the dipolar
effect of $^{52}$Cr is weaker than $^{87}$Rb. The absolute
frequency difference due to dipolar interaction, however, is
larger in $^{52}$Cr because of the faster spin dynamics involved
due to the large spin exchange term in $^{52}$Cr.

In summary, we have studied dipolar effects in spin mixing between
two atoms trapped in a single optical lattice site.
While this effect is small, and can be ignored completely
for spherical harmonic traps, we find it
is observable inside cylindrical harmonic traps,
especially for oblate shaped traps with $\lambda>3$.
We hope this study will
stimulate experimental efforts
aimed at observing dipolar effect in spin mixing.

We thank Drs. K. Bongs, T. Pfau, L. Santos, and K. Sengstock for enlightening
discussions. This work is supported by CNSF and NSF.


\begin{thebibliography}{99}
\bibitem{note}
We have also compared these variational calculations
with a more complete approach using an
exact solution basis for two atoms interacting through a contact
pseudo-potential inside a harmonic trap \cite{Wilken,cd}.
\end{thebibliography}
\end{document}